\documentclass[%
reprint,
 amsmath,amssymb,
 aps,
]{revtex4-2}

\usepackage{graphicx}
\usepackage{dcolumn}
\usepackage{bm}
\usepackage{amsmath,accents,amssymb}
\usepackage{subcaption}
\usepackage{graphicx}
\usepackage{placeins}
\usepackage{comment}
\usepackage{xcolor}
\usepackage{booktabs}

\def\Ka{{\rm Ka}}
\def\Ca{{\rm Ca}}
\def\We{{\rm \Ca^{-1}}}
\def\R{{\rm Re}}

\usepackage{xparse}
\ExplSyntaxOn
\NewDocumentCommand{\mref}{m}{\quinn_mref:n {#1}}
\seq_new:N \l_quinn_mref_seq
\cs_new:Npn \quinn_mref:n #1
 {
  \seq_set_split:Nnn \l_quinn_mref_seq { , } { #1 }
  \seq_pop_right:NN \l_quinn_mref_seq \l_tmpa_tl
  ( 
  \seq_map_inline:Nn \l_quinn_mref_seq
    { \ref{##1},\nobreakspace } 
  \exp_args:NV \ref \l_tmpa_tl 
  ) 
 }
\ExplSyntaxOff

\begin{document}

\preprint{APS/123-QED}

\title{Absolute and convective instabilities in a liquid film \\over a substrate moving against gravity}

\author{Fabio Pino}
 \email{fabio.pino@vki.ac.be}
\affiliation{The von Karman Institute for Fluid Dynamics, EA Department, Sint Genesius Rode, Belgium}
\affiliation{Transfers, Interfaces and Processes (TIPs), Université libre de Bruxelles, 1050 Brussels, Belgium}
\author{Miguel A. Mendez}%
\affiliation{The von Karman Institute for Fluid Dynamics, EA Department, Sint Genesius Rode, Belgium}%
\author{Benoit Scheid}%
\affiliation{Transfers, Interfaces and Processes (TIPs), Université libre de Bruxelles, 1050 Brussels, Belgium}%

\date{\today}

\begin{abstract}
The drag-out problem for small Reynolds numbers ($\R$) admits the Landau-Levich-Derjaguin (LLD) solution for small capillary numbers ($\Ca$), and Derjaguin's solution for large $\Ca$. We investigate whether these solutions are absolutely or convectively unstable, solving the Orr-Sommerfeld eigenvalue problem. For Derjaguin's solution, we show that the LLD solution is convectively unstable for $\Ka<17$ and absolutely unstable for $\Ka \gtrsim 0.15 \,\R^{1.7}$ for $\R > 10$. For water ($\Ka=3400$), the LLD solution is always convectively unstable. The absolute instability is observed only when the dip-coated film is additionally fed from above.
\end{abstract}

\maketitle

\section{Introduction}\label{sec1}

The drag-out problem consists of estimating the thickness of the liquid clinging to a substrate withdrawn from a bath at constant velocity $U_p$ \citep{landau1988dragging,deriagin1964film,derjaguin1993thickness}. Its solution is central in the dip-coating process, where a protective liquid layer is deposited on a substrate \citep{weinstein2004coating}.
In most applications, the smoothness of the deposited layer is the principal criterion for quality. Understanding and predicting the evolution of free surface instabilities could ease control and sensing tasks \citep{pier2003open,huerre1985absolute}. 

For falling films over inclined plates, experiments \citep{liu1993measurements}, analytical and numerical analysis \citep{brevdo1999linear} proved that the so-called hydrodynamic (or Kapitza) instabilities are always convective \citep{kalliadasis2011falling}. For a liquid film suspended under a horizontal plate, instabilities are absolute \citep{sterman2017rayleigh}. Between these extremes, a critical plate inclination angle separating absolute and convective regimes has been identified with experiment \citep{brun2015rayleigh}, simplified model, Direct Numerical Simulations (DNS) \citep{scheid2016critical} and linear stability of the Navier-Stokes equations \citep{pino2024absolute} for a large range of Reynolds numbers.

Similarly, in the case of a liquid film over a substrate moving against gravity, perturbations on a sufficiently thin film should be convected upwards, while those in a sufficiently thick film should be convected downwards, like in falling films. There must, therefore, be a window of intermediate thicknesses for which perturbations might propagate upwards and downwards, reminiscent of absolute instability. Determining whether instability is absolute or convective can be crucial for industrial applications.

To the author's knowledge, the existence of a convective to absolute threshold in a vertical liquid film over a moving substrate has never been explored. In this work, we identify this threshold for a wide range of liquid properties and operating conditions by solving the Orr-Sommerfeld eigenvalue problem for a 2D liquid film.

\begin{figure*}
\centering
  \begin{subfigure}[b]{0.24\textwidth}
  \centering
    \includegraphics[width=\textwidth]{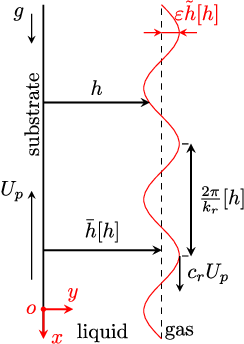}
    \caption{}
  \end{subfigure}
  \hfill
  \begin{subfigure}[b]{0.24\textwidth}
  \centering
    \includegraphics[width=0.65\textwidth]{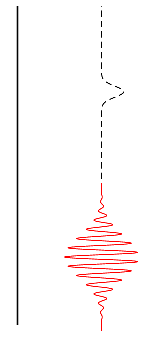}
    \caption{}
  \end{subfigure}
  \begin{subfigure}[b]{0.24\textwidth}
  \centering
    \includegraphics[width=0.55\textwidth]{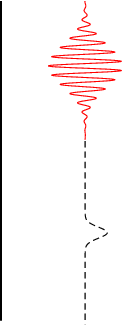}
    \caption{}
  \end{subfigure}
  \hfill
  \begin{subfigure}[b]{0.24\textwidth}
  \centering
    \includegraphics[width=0.65\textwidth]{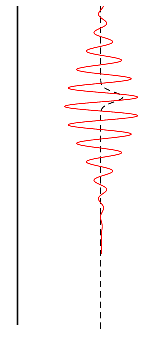}
    \caption{}
  \end{subfigure}
  \caption{(a) Schematic of the investigated configuration: a liquid film over a substrate moving with velocity $U_p$, with thickness $h$ decomposed into a base state $\bar h [h]$ and a harmonic perturbation with wavelength $2\pi[h]/k_r$, phase speed $c_r U_p$ and amplitude $\varepsilon \tilde h [h]$ with $\Tilde{h}, \bar h, k_r, c_r\sim O(1)$, $[h]$ the length scale (see text) and $\varepsilon \ll 1$. The fate (red continuous line) of an initial pulse (dashed black line) is also illustrated in (b) convectively downward, (c) convectively upward, and (d) absolutely unstable flat film solution.}
  \label{fig:problem_description} 
\end{figure*}


\section{Problem Set and Film Regimes}
\label{sec:problem_des}

The problem set is illustrated in Figure~\ref{fig:problem_description} with the relevant parameters. A liquid film of thickness $h(x,t)$ evolves on a vertical substrate rising at a velocity $U_p$. The liquid has density $\rho$ dynamic viscosity $\mu$, kinematic viscosity $\nu=\mu/\rho$ and surface tension at the gas-liquid interface $\sigma$.

Depending on the liquid properties and the velocity $U_p$, different liquid film regimes exist \citep{groenveld1970laminar}. In dimensionless form, the film thickness $\hat{h}=h/[h]$ with $[h]=(\mu U_p/\rho g)^{1/2}$ depends on the Reynolds number $\R=U_p[h]/\nu=(U_p^3/ g\nu)^{1/2}$, and the capillary number $\Ca=\mu U_p/\sigma$. These numbers can be linked by the Kapitza number $\Ka= \R^{2/3}/\Ca=\sigma/(\rho g^{1/3} \nu^{4/3})$ which only dependents on the fluid properties.

In the limit of $\Ca\rightarrow 0$ and perfectly wetting conditions, the Landau-Levich-Derjaguin (LLD) solution gives the steady flat film thickness  ($\hat{h}=\overline{h}$) given by $\overline{h}=0.9458\Ca^{1/6}$ \citep{derjaguin_thickdipcoating,landau1988dragging}. An extension of the LLD solution to account for gravity was found by Wilson \citep{wilson1982drag} who obtained $\bar{h}=0.9458\Ca^{1/6} - 0.3889\Ca^{1/2}$; this was shown to be valid up to $\Ca\sim O(10^{-1})$\citep{krechetnikov2006surfactant}.
For the same $U_p$, \citet{snoeijer_plate_withdrawn} found a one-parameter family of film solutions in the range $\bar{h}\in[\sqrt{3}-0.688\Ca^{1/6},\sqrt{3}]$, appearing for partially wetting fluids just above the critical substrate speed at which the film starts to cling \citep{snoeijer2006avoided}. Note that $\bar h=\sqrt{3}$ corresponds to the transition at which the mean flow rate changes sign and is thus directed downwards, as for falling films \citep{kalliadasis2011falling}.

In the limit of $\Ca\rightarrow \infty$, several authors have shown that $\bar{h}$ reaches a plateau, in line with the maximum flux theory from \citet{derjaguin1993thickness}, which predicts $\bar{h} = 1$. \citet{groenveld1970laminar} derived the upper limit $\bar{h} = 0.66$, while experimental data in \cite{spiers1974free} shows that this should be at $\bar{h} \approx 0.8$. In this work, we determine if these solutions for small and large $\Ca$ are absolutely or convectively unstable. We compute the threshold between absolute and convective instability in the parameter space $\R-\Ka$ for $\bar{h}=1$, and in the space $\bar{h}-\R$ for $\Ka=3400$ (water) and $\Ka=4$ (corn oil).

\section{The Orr-Sommerfeld Eigenvalue Problem}

We define a nondimensional reference system (\textit{O}$\hat{x}\hat{y}$) with $\hat{x}=x/[h]$ aligned with the flow direction, pointing downwards towards the bath, $\hat{y}=y/[h]$ along the wall-normal direction, pointing towards the free-surface. In general, we scale a variable $a$ by a reference quantity $[a]$ to obtain its dimensionless counterpart $\hat{a}=a/[a]$. The over-bar $\Bar{a}$ denotes the nondimensional variable in the stationary solution, and the tilde $\tilde{\cdot}$ denotes the small perturbations at $O(1)$.

Table \ref{table_Dim} provides the definition and the resulting expression for the reference quantities used to derive the dimensionless problem in this work. The liquid film dynamics is governed by the viscous 2D Navier-Stokes equations, with a velocity vector $\mathbf{v}=(u(x,y),v(x,y))$ and a pressure field $p(x,y)$, coupled with the non-slip condition at the substrate ($\hat{y}=0$), which sets $\mathbf{v}(0)=(-U_p,0)$ and thus $\hat{\mathbf{v}}(0)=(-1,0)$, and a set of kinematic and dynamic boundary conditions at the free surface ($\hat{y}=\hat{h}$), which account for the interface continuity and the force balance (see \cite[Chapter~2]{kalliadasis2011falling} and \cite{mendez2021dynamics}). The steady-state solution (base state) is given by a flat interface ($\hat{h}=\overline{h}$) and a parallel shear flow:
\begin{equation}
\label{eq:base_state}
\bar{u}(\hat{y}) = -\frac{1}{2}\hat{y}^2 + \Bar{h}\hat{y} - 1,\qquad
\bar{v}(\hat{y}) = 0,\qquad
\bar{p}(\hat{y}) = 0.
\end{equation}

Integrating \eqref{eq:base_state} across the $\hat{y}$ direction provides the dimensionless flow rate per unit width in stationary conditions, linked to $\bar{h}$ by a cubic law:
\begin{equation}
    \Bar{q} = \frac{1}{3}\Bar{h}^3 - \Bar{h},
\end{equation} where it is worth noticing that the reference flow rate per unit width is $[q]=[u][h]$ and that $\Bar{h}=\sqrt{3}$ identifies the threshold between the drag-out ($\Bar{q}<0$) and the falling film ($\Bar{q}>0$) regime.

\begin{table}[]
\begin{tabular}{ccc}
\begin{tabular}[c]{@{}c@{}}Reference Quantity\end{tabular} & Definition & Expression \\ \toprule
$[h]$ & $(\mu [u]/\rho g)^{1/2}$ & $(\mu U_p/\rho g)^{1/2}$ \\
$[u]\quad\text{and}\quad[v]$ &  $U_p$ & $U_p$ \\
$[p]$  &  $\rho g [h]$ &  $\rho g(\mu U_p/\rho g)^{1/2}$          \\
$[t]$ & $[x]/[u]$ & $(\mu U_p/\rho g)^{1/2}/U_p$   \\
$[q]$ & $[h][u]$ & $(\mu U_p^3/\rho g)^{1/2}$
\end{tabular}
\caption{Reference quantities.}
\label{table_Dim}
\end{table}

To analyse the development of primary instabilities around the base state, the dependent variables are decomposed into a base state and a small perturbation:
\begin{equation}
\label{eq:exp_variab}
    \hat{u} = \bar{u} + \varepsilon\tilde{u},\quad \hat{v}= \bar{v}  + \varepsilon\tilde{v},\quad \hat{p} = \bar{p}  + \varepsilon\tilde{p},\quad \hat{h} = \bar{h}  + \varepsilon\tilde{h},
\end{equation}
where $\varepsilon\ll1$ is a small parameter. Injecting these new variables into the governing equations and collecting the term at \textit{O}$(\varepsilon)$, yields the linearized perturbation (Navier-Stokes) equations:
\begin{subequations}\label{eq:linearized_eqs}
\begin{gather} 
    \partial_{\hat{x}}\tilde{u} + \partial_{\hat{y}}\tilde{v} = 0\label{linearize_eq:1}, \\
    \R\Big(\partial_{\hat{t}}\tilde{u} + \Bar{u}\partial_{\hat{x}}\tilde{u} + \tilde{v} D\Bar{u}\Big) = -\partial_{\hat{x}}\tilde{p} + \nabla^2\tilde{u}\label{linearize_eq:2},\\
    \R\Big(\partial_{\hat{t}}\tilde{v} + \Bar{u}\partial_{\hat{x}}\tilde{v}\Big) = -\partial_{\hat{y}}\tilde{p} + \nabla^2\tilde{v}\label{linearize_eq:3},
\end{gather}   
\end{subequations}
where $D(\cdot)=\partial_{\hat{y}}(\cdot)$ is the wall-normal differential operator and with boundary conditions at the substrate $\hat{y}=0$:
\begin{equation}
    \tilde{u}=\tilde{v}=0,\label{linearize_eq:bc1}
\end{equation}
and at the free surface $\hat{y}=\hat{h}$:
\begin{subequations}
\begin{gather}
\tilde{v} = \partial_{\hat{t}}\tilde{h} + \Bar{u}\partial_{\hat{x}}\tilde{u}\label{linearize_eq:bc2},\\
\tilde{p} = 2\partial_{\hat{y}}\tilde{v} - \We\partial_{\hat{x}\hat{x}}\tilde{h}\label{linearize_eq:bc3},\\
\tilde{h} = \partial_{\hat{y}}\tilde{u} + \partial_{\hat{x}}\tilde{v}\label{linearize_eq:bc4}.
\end{gather}   
\end{subequations}

The linearized equations \eqref{eq:linearized_eqs} are recast to eliminate the streamwise velocity perturbation $\tilde{u}$. Taking the divergence of the linearized momentum conservation equations ($\partial_{\hat{x}}$\eqref{linearize_eq:2}+$\partial_{\hat{y}}$\eqref{linearize_eq:3}) and using the continuity equation \eqref{linearize_eq:1}, yields the following equation:
\begin{equation}
\label{eq:pressure_laplacian_1}
    \nabla^2\tilde{p}=-2\R D\bar{u}\partial_{\hat{x}}\tilde{v}\,.
\end{equation}

Differentiating \eqref{linearize_eq:3} with respect to $\hat{y}$ and evaluating it at the unperturbed free surface $\bar{h}$ and noting that $D\bar{u}|_{\bar{h}}=0$, yields: 
\begin{equation}
\begin{split}
\label{eq:equation_local_1_1}
    -\partial_{\hat{y}\hat{y}}\tilde{p}|_{\bar{h}} =\R (\partial_{\hat{t}\hat{x}}\tilde{v}|_{\bar{h}} + \bar{u}|_{\bar{h}}\partial_{\hat{x}\hat{y}}\tilde{v}|_{\bar{h}}) - \partial_{\hat{y}}\nabla^2\tilde{v}|_{\bar{h}}.
\end{split}
\end{equation}
Injecting \eqref{eq:equation_local_1_1} in \eqref{eq:pressure_laplacian_1}, we deduce that:
\begin{equation}
\label{eq:eq_99}
\begin{split}
\partial_{\hat{x}\hat{x}}\tilde{p}|_{\bar{h}}& = -\partial_{\hat{y}\hat{y}}\tilde{p}|_{\bar{h}} =\\&= \R (\partial_{\hat{t}\hat{x}}\tilde{v}|_{\bar{h}} + \bar{u}|_{\bar{h}}\partial_{\hat{x}\hat{y}}\tilde{v}|_{\bar{h}}) - \partial_{\hat{y}}\nabla^2\tilde{v}|_{\bar{h}}\,.
\end{split}
\end{equation}

By differentiating \eqref{linearize_eq:bc3} with respect to $\partial_{\hat{x}\hat{x}}$ and equating it with \eqref{eq:eq_99} gives:
\begin{equation}
\begin{split}
\label{eq:equation_local_2_1}
    3\partial_{\hat{x}\hat{x}\hat{y}}\tilde{v}|_{\bar{h}}& - \We\partial_{\hat{x}\hat{x}\hat{x}\hat{x}}\tilde{h} \\&+ \partial_{\hat{y}\hat{y}\hat{y}}\tilde{v}|_{\bar{h}} - \R(\partial_{\hat{t}\hat{x}}\tilde{v}|_{\bar{h}} + \bar{u}|_{\bar{h}}\partial_{\hat{x}\hat{y}}\tilde{v}|_{\bar{h}})=0\,.
\end{split}
\end{equation}

Taking the Laplacian of the wall-normal linearized momentum conservations equation ($\partial_{\hat{x}\hat{x}}$\eqref{linearize_eq:3} + $\partial_{\hat{y}\hat{y}}$\eqref{linearize_eq:3}) and  using \eqref{eq:pressure_laplacian_1} leads to the equation: 
\begin{equation}
\label{eq:eq_333_1}
    \nabla^2(\R \,\partial_{\hat{t}}\tilde{v} - \nabla^2\tilde{v}) + \R(1 + \bar{u}\nabla^2)\partial_{\hat{x}}\tilde{v}=0.
\end{equation}

The streamwise derivative of the tangential stress balance at the unperturbed interface ($\partial_{\hat{x}}$\eqref{linearize_eq:bc4}), considering the continuity equation \eqref{linearize_eq:1}, gives:
\begin{equation}
    \partial_{\hat{x}}\tilde{h} - \partial_{\hat{x}\hat{x}}\tilde{v} + \partial_{\hat{y}\hat{y}}\tilde{v} = 0\,.
\end{equation}

The linearized equations are further simplified by introducing the stream function $\Psi$ reading:
\begin{equation}
\tilde{u} = \partial_{\hat{y}}\Psi, \qquad\qquad \tilde{v} = -\partial_{\hat{x}}\Psi,
\end{equation}
and assuming a normal mode solution of the form:
\begin{equation}
\label{eq:def_normal_mod_strf_1}
\begin{split}
    \Psi = \frac{1}{2}\varphi(\hat{y}) \exp(i(k\hat{x} - \omega\hat{t})) + \text{c.c.}, \\\tilde{h} = \frac{1}{2}\eta\exp(i(k\hat{x} - \omega\hat{t})) + \text{c.c.},
\end{split}
\end{equation}where c.c. stands for complex conjugate, $\varphi(\hat{y}) = \varphi_r(\hat{y}) + i\varphi_i(\hat{y})$ is the amplitude, $k=k_r + ik_i$ is the wavenumber, $\omega = \omega_r + i\omega_i$ the angular frequency and $c=c_r + ic_i=\omega/k$ the phase speed of the perturbation. Injecting \eqref{eq:def_normal_mod_strf_1} in the governing equation \eqref{eq:eq_333_1} and in the boundary conditions \mref{linearize_eq:bc1,linearize_eq:bc2,eq:equation_local_1_1,eq:equation_local_2_1} yields the following Orr-Sommerfeld eigenvalue problem in terms of the operators $\mathbf{A}$ and $\mathbf{B}$: 
\begin{equation}
    \label{eq:Orr_Sommerfeld}
\begin{gathered}    
    \mathbf{OS}(k,\omega,Re)\varphi(\hat{y}) =\\= [\mathbf{A}(k,Re) - c\mathbf{B}(k,Re)]\varphi(\hat{y})=0,   
\end{gathered}
\end{equation}
where $\mathbf{A}(k,\R)\varphi(\hat{y})$ is given by:
\begin{equation}
\label{eq:Orr_Sommerfeld_A}
    \mathbf{A}(k,\R) = (D^2-k^2)^2\varphi(\hat{y}) - i\R k[\Bar{u}(D^2-k^2)+1]
\end{equation}
and $\mathbf{B}=-\partial_c\mathbf{OS}$ is given by:
\begin{equation}
\label{eq:Orr_Sommerfeld_B}
    \mathbf{B}(k,Re) = -i\R k(D^2 -k^2),
\end{equation}
and with boundary conditions $\mathbf{OS}_{\rm BC}\varphi|_{0,\hat{h}}=0$ defined as:
\begin{subequations}
\label{eq:bcs}
\begin{gather}
\varphi(0)=D\varphi(0)= 0,\\
\eta = \varphi(\hat{h})/(c-a),\label{kinematic_con_eq}\\
\begin{split}
[(D^2 - 3k^2) + &i \,\R \,k (c - d)]D\varphi(\hat{h})+\\ -& i\eta \We \,k^3 = 0,\label{eq:normal_stress_cond}    
\end{split}\\
(D^2 + k^2)\varphi(\hat{h}) - \eta = 0, \label{eq:bc_OS_4}
\end{gather}
\end{subequations}
where $d = \Bar{u}(\Bar{h}) = (\Bar{h}^2/2-1)$ is the base-state velocity at the air-liquid interface.

\section{Results}

\begin{figure*}
  \begin{subfigure}{0.49\textwidth}
      \centering
    \includegraphics[width=\linewidth]{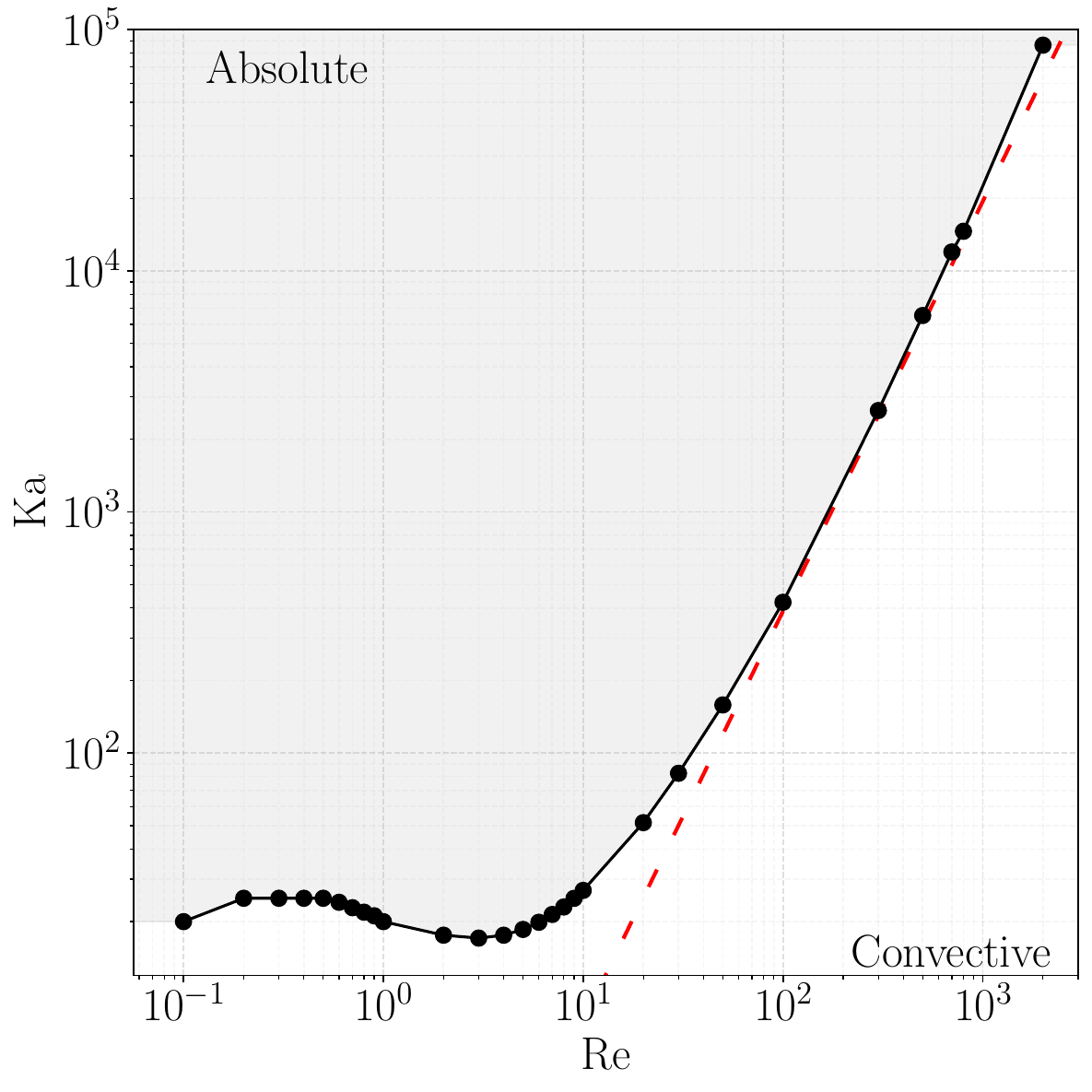}
    \caption{}
    \label{}
  \end{subfigure}
  \hfill
  \begin{subfigure}{0.5\textwidth}
    \centering
    \includegraphics[width=\linewidth]{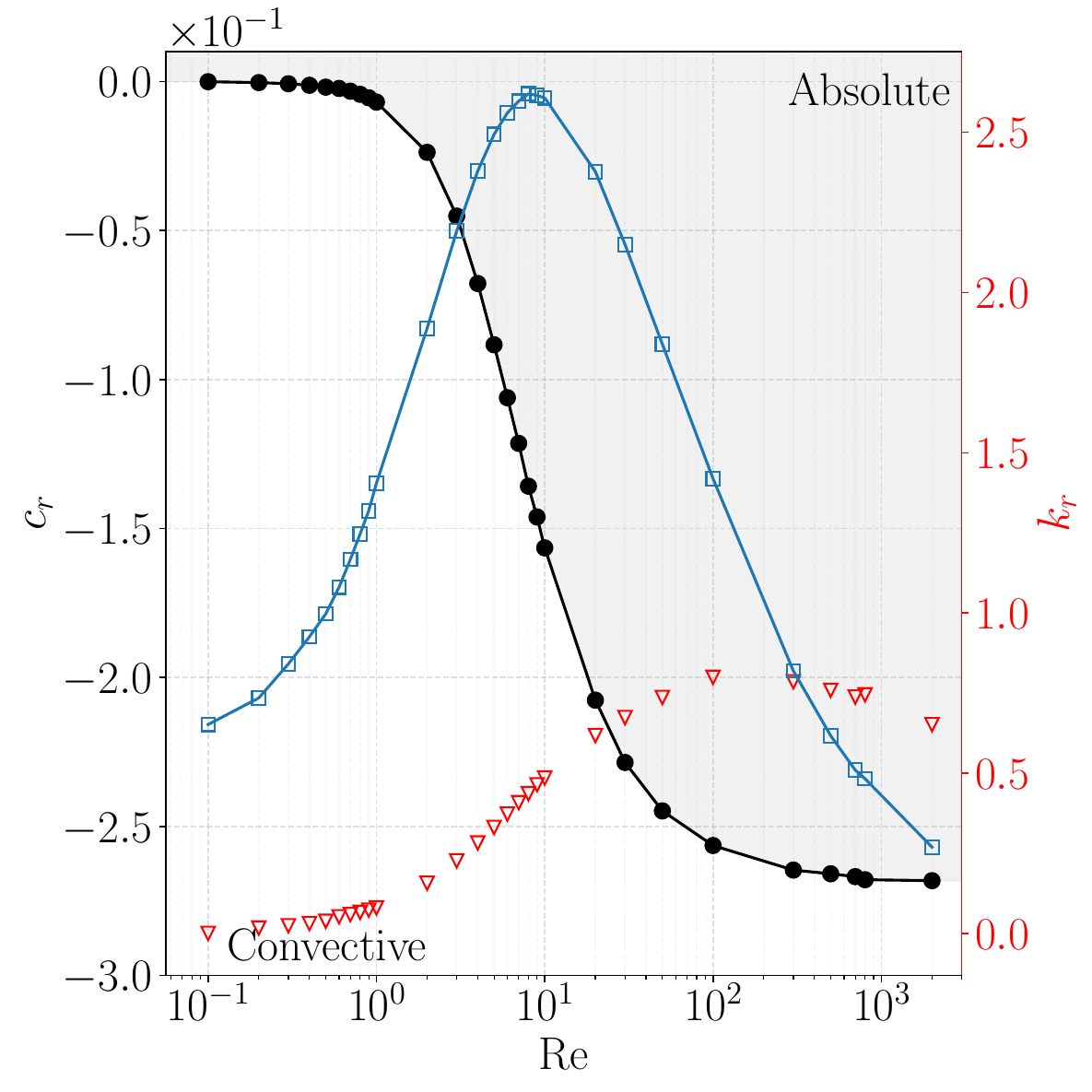}
    \caption{}
    \label{}
  \end{subfigure}
    \caption{(a) Absolute/convective threshold (black dots) for solutions with $\bar h=1$ in the $\Ka-\R$ space, and (b) corresponding spatial phase speed $c_r$, real wavenumber $k_r$ (red triangles), and nondimensional capillary wavenumber $2\pi/\hat{\ell}_{c}$ (blue squares), as a function of $\R$.}
    \label{fig:ACvaryingKa}
\end{figure*}

The absolute-convective threshold of the solution $\Bar{h}=1$ presents two regimes, as shown in Figure~\ref{fig:ACvaryingKa}a. 
For $\Ka<30$, there is a minimum of $\Ka=17$ at $\R=3$ and a maximum of $\Ka=25$ at $\R=0.4$. 
This result shows that for fluids with $\Ka<17$, the solution is unconditionally convectively unstable for large $\Ca$ regardless of the Reynolds number. For $\R \lesssim 10$ the instability always becomes absolute for $\Ka>25$, whereas for $\R > 10$, the convective to absolute threshold reaches a trend which goes as $\Ka\approx\,0.15\R^{1.7}$  or, in terms of capillary number, as $\Ca^{-1}\approx \,0.15\R$. 

The phase speed $c_r$ of the neutral modes at the threshold is always negative, as shown in Figure~\ref{fig:ACvaryingKa}b. The unstable perturbations are entrained by the substrate, with waves going 
upwards. This phase speed is maximum at zero for $\R\rightarrow 0$ and reaches a plateau at $c_r\approx-2.7\times 10^{-1}$ for $\R \gtrsim 10^3$. These neutral modes have small wave numbers for all $\R$. The wavenumber increases with $\R$ reaching a peak of $k_r\approx0.8$ around $\R\approx 100$. The increase of $k_r$ with $\R$ suggests a predominant role of surface tension, which is corroborated as it becomes larger than the capillary wavenumber $2\pi/\hat{\ell}_{c}=2\pi\R^{1/3}/\Ka^{1/2}$ (blue line with squares), where $\hat{\ell}_{c}=\ell_{c}/[h]$ with $\ell_c=\sqrt{\sigma/(\rho g)}$ the capillary length. It means that for perturbation wavenumbers larger than the capillary wavenumber, surface tension predominates over gravity, which is the case for $\R > 300$.

\begin{figure*}
  \begin{subfigure}{0.50\textwidth}
      \centering
    \includegraphics[width=\linewidth]{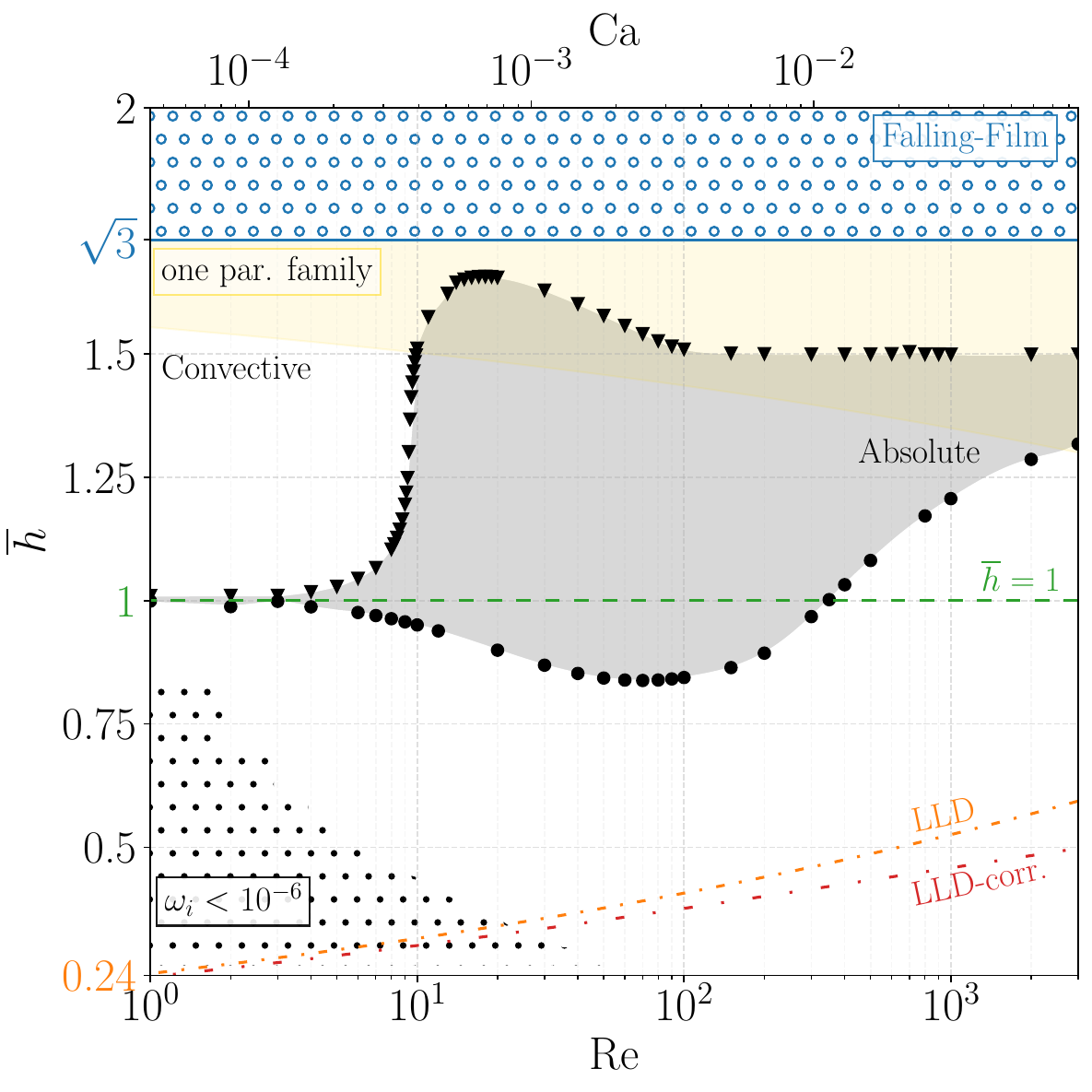}
    \caption{}
    \label{}
  \end{subfigure}
  \hfill
  \begin{subfigure}{0.47\textwidth}
    \centering
    \includegraphics[width=\linewidth]{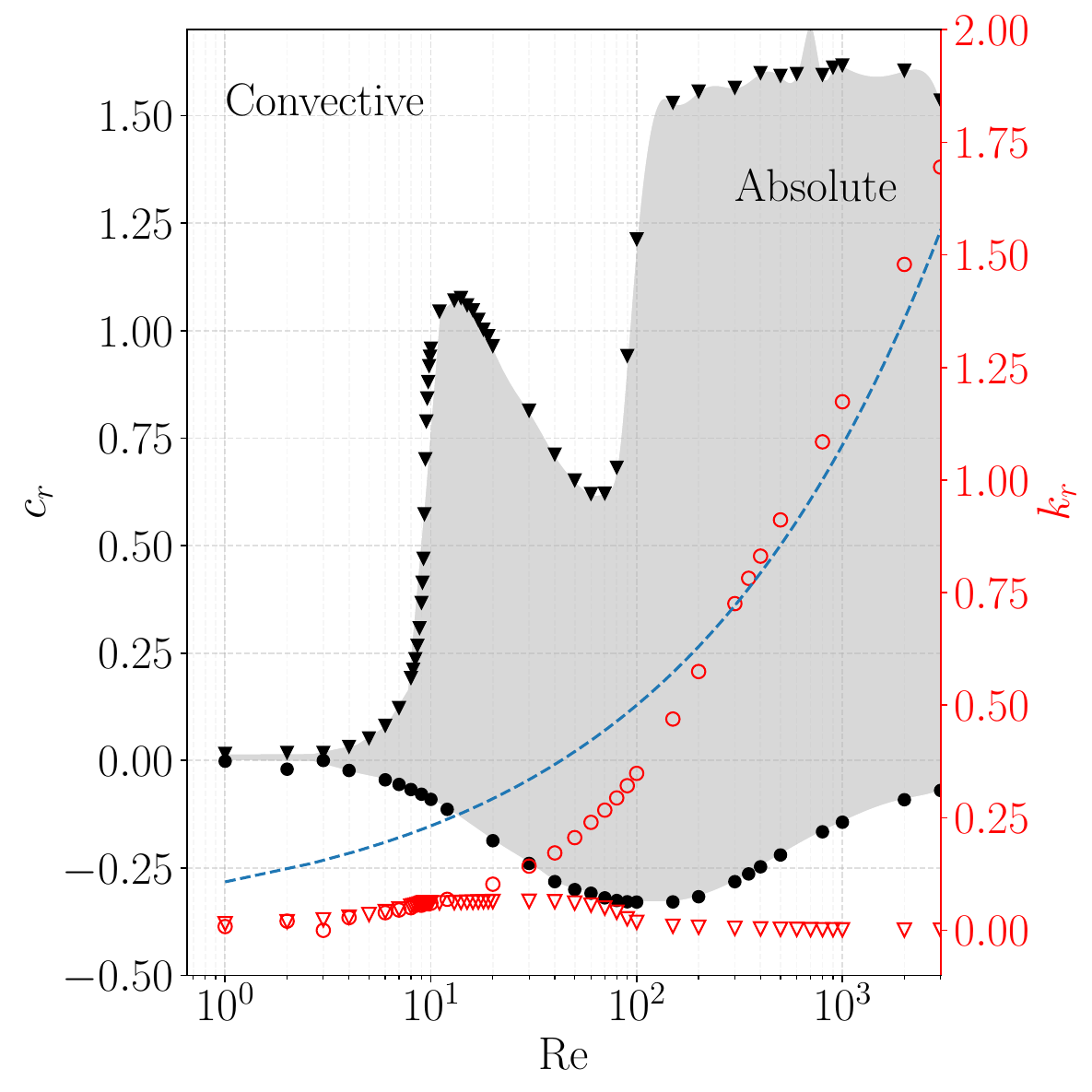}
    \caption{}
    \label{}
  \end{subfigure}
    \caption{(a) Absolute/convective threshold (black symbols) in the $\bar h-\R$ space and, (b) in the $c_r-\R$ space for water ($\Ka=3400$). In (a), the $\Ca$ number is indicated on the top axis, the dotted area corresponds to $\omega_i<10^{-6}$, and the yellow region indicates thick solutions (see text for details). In (b), the spatial wavenumber $k_r$ is shown in red and compared to the nondimensional capillary wavenumber $2\pi/\hat{\ell}_{c}$ (dashed blue line). }
    \label{fig:threshold_water}
\end{figure*}

Figure~\ref{fig:threshold_water} shows the window of absolute instability in the $\bar h-\R$ space for water, bounded by an upper and lower branch of solutions. 
The lower branch (black dots) stems from $\bar h=1$ and develops mainly into the region for $\bar h<1$, with a minimum at $\R\approx 80$. The curve extends into the region for $\R>300$ for $\bar h>1$. In the $c_r-\R$ space in figure~\ref{fig:threshold_water}b, the neutral waves, associated with the lower branch, always travel upwards with a negative phase speed, even for $\bar h>1$. The associate wavenumber (red circles) continuously increases with $\R$, overtaking the $2\pi/\hat{\ell}_{c}$ curve at around $\R\approx 300$, hence showing that surface tension tends to prevail over gravity, as in the $\Ka-\R$ space for $\bar h=1$.

The upper branch (black triangles) also stems from $\bar h=1$ and extends into the region for $\bar h>1$. Through a steep increase, it attains a maximum of $\bar h\approx1.65$ for $\R\approx 10$. A decay follows this extremum until $\R\approx100$, beyond which the curve reaches a plateau at $\bar h\approx1.5$. The phase speed of the neutral mode at the threshold is always positive, and the corresponding wavenumber is always below 0.1, corresponding thus to long wavelengths. The phase speed has a maximum around unity, meaning that waves move at the same speed as the substrate but in the opposite direction. For increasing $\R$ beyond this maximum, the phase speed decays before rising again to reach a plateau at $c_r \approx 1.5$ for $\R\gtrsim 100$. 

In particular, the upper branch for thick films ($\bar h > 1$) delimitates two distinct regimes at around $\R\approx 10$. For smaller $\R$, gravity dominates the wave dynamics, driving the perturbations convectively downwards. For larger $\R$, inertia prevails over gravity, and the perturbations are then also governed by the substrate entrainment, thus propagating absolutely in the whole domain. An increase in surface tension effects accompanies the transition between the two regimes. Interestingly, in the range of small $\R \lesssim 1$, the solution $\bar h = 1$ is the only one being absolutely unstable, giving rise to standing waves, as perturbations in thinner films will propagate upwards and in thicker films downwards, as anticipated in the introduction.

The LLD solution and Wilson's corrected solution are also plotted in Figure~\ref{fig:threshold_water} in with orange dash-dotted and red dash-dotted lines, respectively. We conclude that these solutions are always convectively unstable for water. In addition, the smallness of the growth rate (dotted area), {\it i.e.} $\omega_i <10^{-6}$, for small $\Ca$ indicates that the instability can hardly be observed in practice. On the contrary, \citet{snoeijer_plate_withdrawn}'s family of solutions (yellow shadowed area) can be absolutely unstable for $\R \gtrsim 10$ and $\bar h \gtrsim 1.5$.

\begin{figure*}
  \begin{subfigure}{0.47\textwidth}
      \centering
    \includegraphics[width=\textwidth]{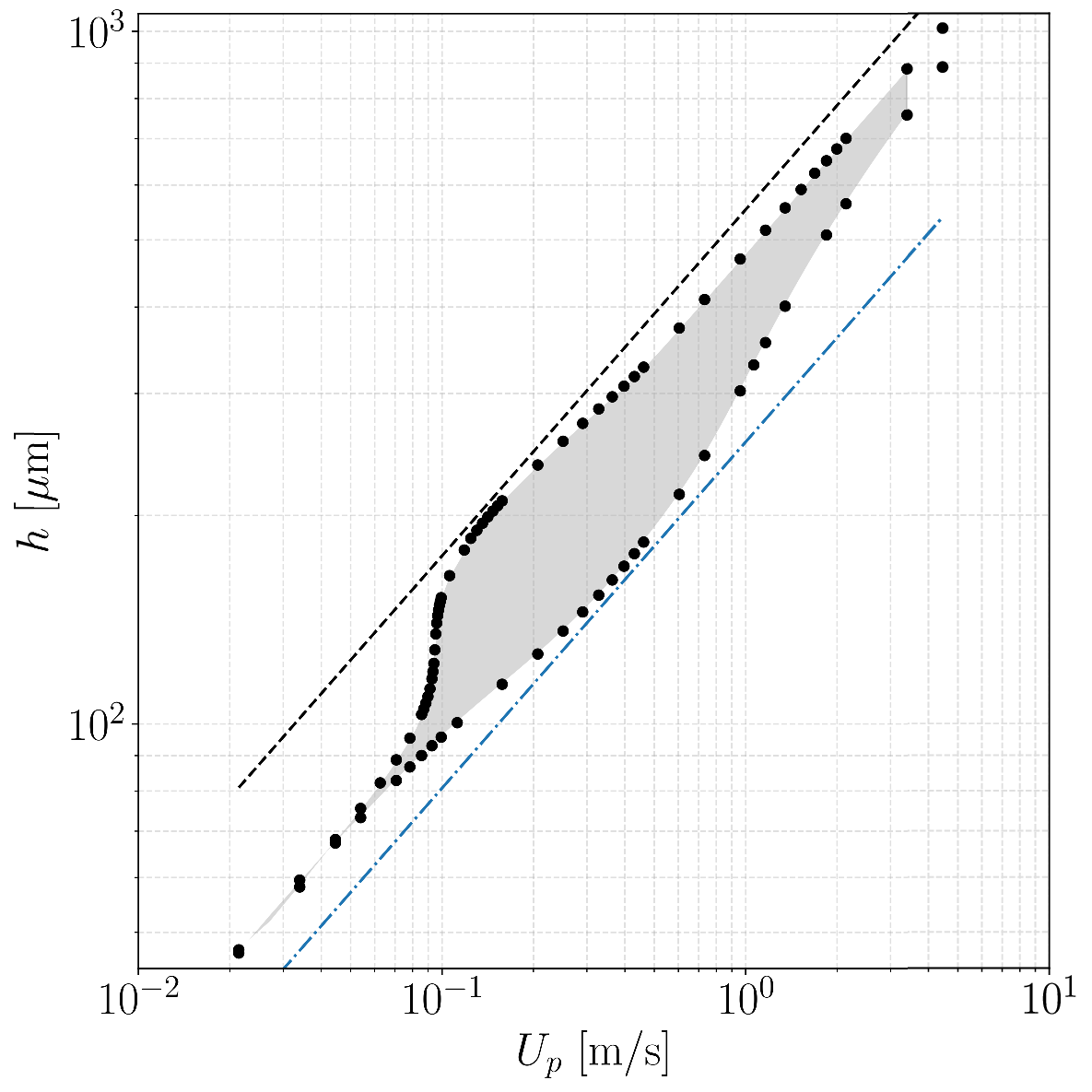}
    \caption{}
    \label{}
  \end{subfigure}
  \hfill
  \begin{subfigure}{0.47\textwidth}
    \centering
    \includegraphics[width=\textwidth]{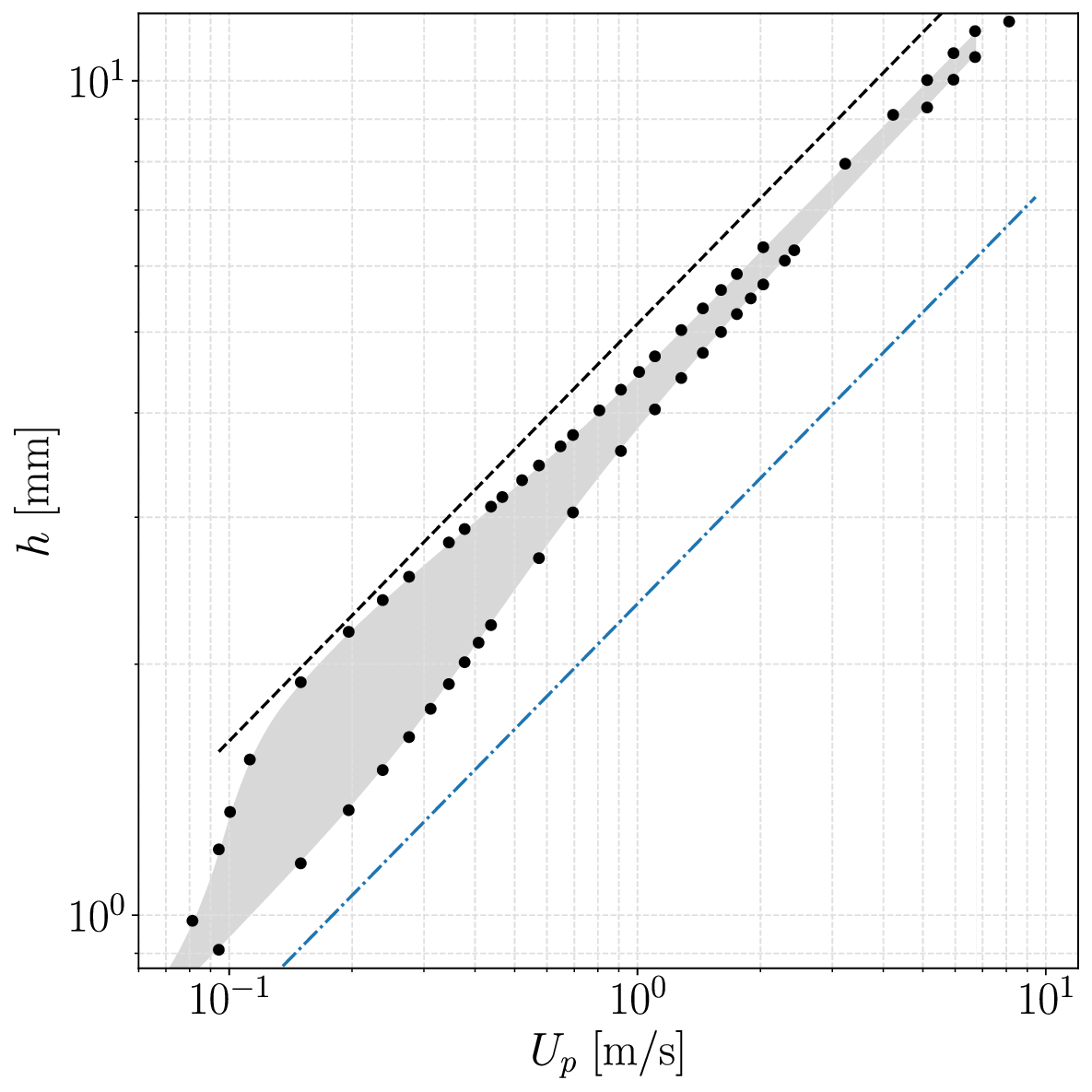}
    \caption{}
    \label{}
  \end{subfigure}
    \caption{Region of absolute instability (grey area) in the dimensional space $h-U_p$ with the boundary between dip-coating and falling film $(\bar h=\sqrt{3})$ (black dashed line) and the maximum thickness attainable in drag out according to \cite{spiers1974free} (blue dash-dotted line) for (a) water ($\Ka=3400$) and (b) corn oil ($\Ka=4$).}
    \label{fig:threshold_dimensional}
\end{figure*}

For illustrative purposes and as a trigger to future experimental work, figure~\ref{fig:threshold_dimensional} locates the area of absolute instability in a dimensional map for (a) water and (b) corn oil, characterized by largely different Kapitza numbers. 
The thickness corresponding to $\bar h=\sqrt{3}$ and $\bar h=0.8$ are also shown in black and blue lines, respectively. The first denotes the limit above which the problem is in the falling film regime, and the second denotes the upper limit, which can be experimentally obtained in the drag-out configuration according to \citet{spiers1974free}. For both liquids, the absolute regime is within these bounds, meaning that the experimental exploration of the window of absolute instability and the associated thresholds to convective instability would require a facility combining drag-out and liquid supply from above.

\section{Conclusions}
Our study first demonstrates that a liquid film over a substrate moving against gravity is always unstable, even though the growth rate in the LLD limit is generally too small for the perturbations to be seen in practice. Second, it demonstrates the existence of a window of absolute instability that should be appealing for further foundational and applied research in coating processes involving moving substrates.

\begin{acknowledgments}
F.P. was supported by a FRIA grant from FNRS. and B.S. is Research Director at F.R.S.-FNRS.
\end{acknowledgments}

\bibliography{Main_text_incl._figures}
\bibliographystyle{apsrev}
\end{document}